\documentclass[english,letterpaper,twocolumn,showpacs,pra,aps]{revtex4}
\usepackage{graphicx}
\usepackage{amssymb}

\makeatletter

\large  

\input epsf

\makeatother

\usepackage{babel}
\makeatother
\begin{document}

\title{Casimir energy of smooth compact surfaces}

\author{Joseph P. Straley$^{1}$ and Eugene B. Kolomeisky$^{2}$}

\affiliation
{$^{1}$Department of Physics and Astronomy, University of Kentucky,
Lexington, Kentucky 40506-0055, USA\\
$^{2}$Department of Physics, University of Virginia, P. O. Box 400714,
Charlottesville, Virginia 22904-4714, USA}

\begin{abstract}
We discuss the formalism of Balian and Duplantier for the calculation of the Casimir energy for an arbitrary smooth compact surface, and use it to give some examples: a finite cylinder with hemispherical caps, the torus, ellipsoid of revolution, a "cube" with rounded corners and edges, and a "drum" made of disks and part of a torus.  We propose a model function which approximately captures the shape dependence of the Casimir energy.
\end{abstract}


\maketitle

\section{Casimir energy}
 
The basis of the Casimir effect is that introduction of a conductor 
into a previously empty space modifies the electromagnetic spectrum, 
and thus the zero-point energy.   We can study this effect using the mode sum generating function
\begin{equation}
\label{defmodesum}
S(\Omega) = \sum_{\alpha}  \exp(-\omega_{\alpha}/\Omega)
\end{equation}
where $\alpha$ labels the modes,  $\omega_{\alpha}$ is a mode frequency 
for the disturbed system, and $\Omega$ is a frequency cutoff.   
Previous work \cite{Weyl} has established that the mode generating 
function for a general spectrum of a three-dimensional system can be 
expanded in powers of $\Omega$ in the form
\begin{equation}
\label{modegenerate}
S(\Omega) = k_{1} (\Omega/c)^{3} {\mathcal V} + k_{2} (\Omega/c)^{2} {\mathcal A} +  (\Omega/c) {\mathcal I} + {\mathcal K} + 2{\mathcal C} c/\Omega
\end{equation}
The leading terms are local quantities:  the contribution is a sum over values 
defined for each point in space.   Specifically, $c$ is the speed of light, $k_{1}$ and $k_{2}$ are pure numbers, $\mathcal V$ is the volume of the 
region, $\mathcal A$ is the surface area of the object, $\mathcal I$ is the surface 
integral of a linear combination of the surface curvatures, and $\mathcal K$ (the Kac number) 
is the surface integral of a quadratic combination of the surface curvatures.   
In the last term $\mathcal C$ (the Casimir term) is a coefficient that depends inversely on a length scale that characterizes the size of the object.  It is shape dependent in a way that has not been well characterized.

The specific problem we study is the change $\delta S$ caused 
by the introduction of a thin conducting surface into an initially empty space.   
For this problem the expansion (\ref{modegenerate}) considerably simplifies:  for empty space we have only the leading term, and since the surface does not 
change the volume in which the modes exist, the first term cancels out in 
the difference;  the third term vanishes because the surface integral is over both 
the inside and outside of the same surface, with oppositely directed surface elements.   
For electromagnetism the second term vanishes too:  in the cases that have been 
treated previously \cite{Weyl} this occurs because the "transverse magnetic" and "transverse electric" 
modes make opposite contributions, but Balian and Duplantier \cite{BD} (BD) have given 
an argument that indicates this is a general property.    The result is that the limit 
$\Omega \rightarrow \infty$ can be taken.   We define the Kac number \cite{kac}
\begin{equation}
\label{kacnumber}
{\mathcal K} = \delta S(\Omega \rightarrow \infty) 
\end{equation}
which represents the number of modes that are created by the introduction of the object, and the Casimir term 
\begin{equation}
\label{casimirnumber}
{\mathcal C}  =  \frac {1}{2c}\lim_{\Omega \rightarrow \infty}    \Omega^{2}   d\delta S/d\Omega
 \simeq   \frac {1}{2c} \sum_{\alpha} (\omega_{\alpha} - {\bar \omega}_{\alpha})
\end{equation}
where $\bar \omega_\alpha$ are the mode frequencies before the introduction of the conductor.
Thus the change in the zero-point energy is $\delta {\mathcal E} = \hbar c{\mathcal C}$. 
The Casimir term is nonlocal (i.e. it cannot be calculated as 
a sum of contributions from separate parts);  it has hitherto been known only for 
a few special geometries (sphere \cite{sphere}, circular cylinder \cite{cylinder}, and cylinder of 
elliptic cross-section \cite{SWK}).   Balian and Duplantier (BD) \cite{BD} have given an approach to the Casimir problem that allows calculation of $\mathcal K$ and $\mathcal C$ for general closed conducting surfaces 
with bounded curvature.  In this paper we will comment on their method, and use it to enrich the literature with some examples.

\section{Balian-Duplantier expansion}

\subsection{Mode expansion in terms of a Green function} 

For case of a fixed frequency, Maxwell's equations can be reduced to the inhomogeneous Helmholtz equation
\begin{equation}
\label{Maxwell1}
(\nabla^{2} + \omega^{2}/c^{2}) \vec B(\vec R) = -\mu_{0} \vec \nabla \times \vec J(\vec R)
\end{equation}
subject to the constraint that the field be solenoidal
\begin{equation}
\vec \nabla \cdot \vec B(\vec R) = 0
\end{equation}
for all $\vec R$ in the region.
The mode frequencies $\omega_{\alpha}$ for a finite region enclosed by a perfect conductor are those that allow a solution to these equations with no sources inside the region, subject to the boundary conditions
\begin{equation}
\label{boundary}
\hat n \cdot \vec B(\vec r) = 0, \quad \quad \hat n \cdot (\vec \nabla \times \vec B (\vec r) ) = 0
\end{equation}
where $\hat n$ is a unit vector perpendicular to the surface and $\vec r$ is any point on the surface. The corresponding mode fields $\vec B_{\alpha}$ form a complete orthonormal basis for the representation of solenoidal fields within the region.  Then in particular 
\begin{equation}
\label{completeness}
\sum_\alpha \vec B_{\alpha} (\vec R) \otimes \vec B_{\alpha} (\vec r)
= \delta (\vec R - \vec r) {\textbf I}
\end{equation}
where $\textbf I$ is the unit tensor and $\bf \otimes$ denotes the dyadic product.
For general frequency $\omega$ we now consider the matrix 
\begin{equation}
\label{defGamma}
{\bf \Gamma} (\vec R, \vec r) = \sum_\alpha \frac {\omega_{\alpha}^{2}}
{\omega_{\alpha}^{2}-\omega^{2}}
\vec B_\alpha (\vec R) \otimes \vec B_\alpha (\vec r)
\end{equation}
Regarded as a function of $\vec R$, ${\bf \Gamma}$ satisfies the inhomogeneous Helmholtz equation
\begin{equation}
(\nabla^{2} + \omega^{2}/c^{2}) {\bf \Gamma}(\vec R, \vec r) = 
-\mu_{0} \vec \nabla_{R} \times (\vec \nabla_{R} \times {\bf I}\delta (\vec R - \vec r))
\end{equation}
and inherits from $\vec B (\vec R)$ the property $\vec \nabla{R} \cdot {\bf \Gamma}=0$ and the conditions that $\hat n_{R} \cdot {\bf \Gamma}(\vec R, \vec r) = 0$ and $\hat n_{R} \cdot (\vec \nabla_{R} \times {\bf \Gamma}(\vec R, \vec r))=0$ when $\vec R$ is on the boundary.  Then ${\bf \Gamma}(\vec R, \vec r) \cdot \vec m$ is the magnetic field at $\vec R$ caused by a magnetic dipole $\vec m$ at
$\vec r$ in the presence of the conductor: it is the Green function for that problem.  We similarly define ${\bf \Gamma}_{0}$ for the space in the absence of the conductor.

We can see from Eq.(\ref{defGamma}) that there is divergent response to the presence of an oscillating magnetic dipole at the resonant frequencies, as would be expected on physical grounds.  Regarding $\omega$ as a complex-valued variable, the Green function has simple poles at the mode frequencies and is analytic elsewhere.  Integrating $\text{tr} {\mathbf \Gamma}(\vec r, \vec r)$ over all space turns (\ref{defGamma}) to a sum of simple poles with residue $\frac {1}{2} \omega_{\alpha}$. For $\omega$ with a small positive imaginary part this reduces to a set of delta functions $i \frac {1}{2} \pi \omega \delta (\omega - \omega_{\alpha})$ from which the mode sum generating function can be constructed.  The argument can be repeated for the region exterior to the conductor and for the whole space in the absence of the conductor, leading to the definition 
\begin{equation}
\label{defPhi}
\Phi(\omega) = 
\int d^3r \; \text{tr} [{\mathbf \Gamma}(\vec r, \vec r; \omega )
- {\bf \Gamma}_0(\vec r, \vec r; \omega )]
\end{equation} 
where the integral is over all space, so that this contains both the interior and exterior modes, and subtracts out the modes for empty space.  It has the property 
\begin{equation}
\label{ImPhi}
\Im \Phi(\omega + i \epsilon) = \frac {1}{2}\pi \sum_{\alpha} \omega 
[\delta (\omega - \omega_{\alpha})- \delta (\omega - \bar \omega _{\alpha})]
\end{equation}
and thus \cite{note1}
\begin{equation}
\label{nocontour}
\delta S(\Omega) = 
\Im \int_0^{\infty + i \epsilon} 
\frac {2}{\pi \omega} \exp(-\omega/\Omega) 
 \Phi(\omega) d\omega
\end{equation}
which motivates the further discussion of $\bf \Gamma$.

\subsection{Integral equation for the Green function}

In the absence of boundaries the magnetic field due to an oscillating magnetic dipole is given by
\begin{equation}
\label{defGamma0}
{\bf \Gamma}_{0}(\vec R, \vec r)\cdot \vec m
= (\vec m (\vec \nabla_{R} \cdot \vec \nabla_{r}) - \vec \nabla_{R}
(\vec m\cdot \vec \nabla_{r})) G_{0}(\rho)
\end{equation}
where 
\begin{equation}
\label{defrho}
\vec \rho = \vec R - \vec r\:, \quad \rho = |\vec \rho| 
\end{equation}
and
\begin{equation}
\label{defGreen0}
G_{0}(\rho) = \frac {e^{i \omega \rho/c}}{4\pi \rho}
\end{equation}
is the Green function for the scalar Helmholtz equation.  

When there is a conducting surface present, a surface current $\vec j_{ind}(\vec r)$ is induced which gives rise to a new magnetic field $\vec B_{ind}$.  Thus we are led to 
define another Green function ${\bf M}$
\begin{eqnarray}
\label{defM}
\vec B_{ind}(\vec R) &=& \mu_{0} \int_{s} d^{2}r \vec \nabla_{R} \times
G_0(\vec R, \vec r)\vec j_{ind}(\vec r)
\nonumber
\\
&\equiv&
\mu_{0} {\bf M}(\vec R, \vec r) \cdot \vec j_{ind}(\vec r)
\end{eqnarray}
where the integral is over the surface of the conductor, and in the second line we introduce the convention that the "repeated index" $\vec r$ implies an integration over the surface. The surface current completely represents the effect of the conducting surface, so that the total field can be regarded as the combination of the field due to the source
alone and to the field created by the induced currents, as if the fields were propagating in free space:
\begin{equation}
\label{currentview}
{\bf \Gamma}(\vec R,\vec r_{0})\cdot \hat m = {\bf \Gamma}_{0}(\vec R, \vec r_{0})\cdot \hat m
+ {\bf M}(\vec R, \vec r) \cdot \vec j_{ind}(\vec r)
\end{equation}
Note that $\vec j_{ind}$ depends implicitly on the location $\vec r_{0}$ and orientation $hat m$ of the source.

The surface provides perfect screening: when the source is inside, there are no fields outside.  The tangential components of $\vec B$ can be discontinuous across the boundary,  due to the presence of $\vec j_{ind}$. BD observe that we can use this to determine $\vec j_{ind}$ in the form
\begin{equation}
\label{defsurfacecurrent}
\vec j_{ind}(\vec R)= \frac {2}{\mu_{0}}  \hat n (\vec R) \times \vec B(\vec R) .
\end{equation}
Here $\hat n$ is directed outwards when the source is within the surface, and inwards when the source is outside. This gives a local relationship between $\vec j_{ind}(\vec R)$ and $\vec B (\vec R)$.  

The field $\vec B(\vec R)$ is the total field due to all currents except the surface current at $\vec R$ (this exclusion is the origin of the factor of $2$).  It is defined as follows: for $\vec R$ close to the surface (at distance $\epsilon$ from it), we imagine using the first version of (\ref{defM}) to calculate $\vec B_{ind}(\vec R)$, but exclude from the surface integral a small region near $\vec R$ that is significantly larger than $\epsilon$. This part of $B_{ind}$ is continuous as $\vec R$ moves through the surface.  The excluded region contributes a field that is discontinuous at the surface. The surface current (\ref{defsurfacecurrent}) is chosen so that $\vec B_{0}$ and the part of $\vec B_{ind}(\vec R)$ just computed are canceled for $\vec R$ just outside the surface.  In taking the limit that $\vec R$ is on the surface, the size of the "small region" can go to zero; the magnetic field $\vec B(\vec R)$ is after all the result of doing the integral (\ref{defM}) over the whole surface.  

Substituting Eq.(\ref{currentview}) gives an integral equation for $\vec j_{ind}$: 
\begin{equation}
\label{jeq}
\vec j_{ind}(\vec R) = \vec j_s (\vec R) + {\bf K}(\vec R, \vec r) \cdot \vec j_{ind}(\vec r)
\end{equation}
where 
\begin{equation}
\label{defK}
{\bf K}(\vec R, \vec r) = 2 \hat n(\vec R) \times {\bf M}(\vec R, \vec r)
\end{equation}
describes how the magnetic field created by a current at $\vec r$ induces a current at $\vec R$, and
\begin{equation}
\vec j_s (\vec R) = 2 \hat n(\vec R) \times 
 {\bf \Gamma}_{0}(\vec R, \vec r) \cdot \vec m 
\end{equation}
is the part of the surface current that is directly due to the presence of the magnetic dipole $\vec m$ at $\vec r$.

The useful feature of (\ref{jeq}) is that it turns a set of three-dimensional partial differential equations for fields with three components into an integral equation for a two-component vector field living on a two-dimensional surface.  This makes the numerical study of the problem greatly simpler.

\subsection{Series representation of the Green function}

The integral equation (\ref{jeq}) can be given a formal solution by repeatedly substituting it into itself, giving an expansion
\begin{equation}
\label{jexpansion}
\vec j_{ind} = \vec j_s + {\bf K}\cdot \vec j_s + {\bf K}\cdot
{\bf K} \cdot \vec j_s + ... .
\end{equation}
This leads to a corresponding expansion for $\bf \Gamma$
\begin{equation}
\label{Gammaexpansion}
{\bf \Gamma} = {\bf \Gamma}_0 + {\bf M}\cdot \vec j_s + 
{\bf M}\cdot{\bf K} \cdot \vec j_s + ... .
\end{equation}

The usual route to solving the electromagnetism problem we are considering is to find a set of fields satisfying Maxwell's equations away from the boundary, then impose the boundary conditions (\ref{boundary}) that enforce continuity of some components of the field, and only at the end (if ever) consider the discontinuous components that correspond to the surface currents and charges.  BD reverse the process and find the surface currents first.   The usual boundary conditions do not play an explicit role.

In defense of BD's approach, we observe that induction of surface currents is how the conducting surface screens the electromagnetic field, and that it necessarily establishes these currents on a local basis.  BD's approach is the way that nature solves the problem.

To convince oneself that we are studying the same problem, it is useful to consider the case of spherical geometry. Suppose that BD's boundary condition is not finding the correct solution.  Then there is a solution to Maxwell's equations that has vanishing tangential components of the magnetic field at the surface, but nonzero radial magnetic field.  Since $\vec r \cdot \vec B$ is a solution to the scalar wave equation, the surface components determine this quantity everywhere outside, in the form $\sum A_{l,m} j_{l}(\omega r/c) Y_{l,m} (\theta, \phi)$.  It is then readily shown \cite{Jackson} that for general $\omega$ this implies nonzero tangential components of the magnetic field, contrary to our starting assumption.  Although an example is less than a proof, the point is that tangential and normal components of the magnetic field are related to each other, and even in the special case of spherical geometry we cannot have one without the other.

The expansion (\ref{jexpansion}) has the virtue that it turns a problem to be solved (\ref{jeq}) into an expansion to be evaluated: the explicit form of the right-hand side of (\ref{jexpansion}) is known.  Yet if we had been interested in the currents and fields at a particular frequency the expansion would be a terrible idea.  The source emits electromagnetic waves which are scattered on every encounter with the surface, and the mode frequencies emerge from the coherence of many such scatterings, giving a divergent response $\vec j_{ind}$. This would appear in this representation as a consequence of the failure of the expansion to converge.  However, the expansion works fine for the intended application: when we chose the contour for Eq.(\ref{nocontour}) to lie along the imaginary frequency axis (and then closing at infinity), the various Green functions have exponential decrease with distance and the sum is rapidly convergent.  It will turn out that just the first nonvanishing term is a good approximation.

Now returning to the evaluation of (\ref{nocontour}) using (\ref{defPhi}),
the leading term of (\ref{Gammaexpansion}) cancels.  This could be anticipated from (\ref{modegenerate}), because the surface being introduced does not change the volume.  For the remaining terms of (\ref{Gammaexpansion}), BD show that the three-dimensional integral in (\ref{defPhi}) can be done analytically to give
\begin{eqnarray}
\label{defPhi1}
\Phi(\omega) =  \sum_{m=1}^{\infty} \Phi_{m}
\nonumber
\\
\Phi_{m} = \frac {1}{2m} \omega \frac {d}{d\omega}
\text {Tr} {\bf K}(\vec r_{1},\vec r_{m})...{\bf K}(\vec r_{3},\vec r_{2}){\bf K}(\vec r_{2},\vec r_{1})
\end{eqnarray}
where the symbol $Tr$ represents the trace over the resulting product as well as surface integrals over the variables $\vec r_{i}$, all of which are on the surface.  This expression can be interpreted as  a sum over all closed paths of $m$ sites.

The sum resembles the expansion of a logarithm.  A similar expression has appeared in the evaluations of the Casimir energy for geometries that allow separation of variables \cite{sphere,cylinder,SWK}.   

On physical grounds, the poles of $\Phi$ should all be for real $\omega$.  Then the integrand for (\ref{nocontour}) is analytic in the upper half-plane.  This allows us to choose a path of integration that goes along the imaginary $\omega$ axis to large values and then returns to the real axis along an arc at large $|\omega|$.   Writing $\omega = iyc$, we find that $\bf K$ and $\bf M$ have no imaginary part on the $y$ axis.  Combining (\ref{defK}), (\ref{defM}), and (\ref{defGreen0}), the explicit form of $\bf K$ is
\begin{equation}
\label{explicitK}
{\bf K}(\vec r_{2}, \vec r_{1})
= - \frac {e^{-y\rho}(1+y\rho)}{2\pi\rho^{3}} (\vec \rho \otimes \hat n_{2} - {\bf I} \vec \rho \cdot \hat n_{2})
\end{equation}
where $\vec \rho = \vec r_{2} - \vec r_{1}$ and $\hat n_{2}$ is the outward normal to the surface at $\vec r_{2}$. 

Despite the singularity at $\rho = 0$, the integral of this expression is not divergent, because in that limit both $\vec \rho$ and $\vec j_{2}$ are tangent to the surface and thus perpendicular to $\hat n_{2}$;  then $\vec \rho (\hat n \cdot \vec j) - (\hat n \cdot \vec \rho)\vec j $ is vanishing like $\rho^{2}$, leaving an integrable $1/\rho$ singularity.

For every closed path of sites on the surface, there is a corresponding one in which the sites are traversed in reversed order.  Physically we might expect these paths to have related values, but the dependence of $\bf K$ on $\vec \rho$ suggests that reversing the path will reverse the sign of each factor in (\ref{defPhi1}), leading to a cancellation when there is an odd number of sites.  For this reason only the even $m$ terms need be considered.  Since $\bf K$ is not symmetric under exchange of labels, the perfect cancellation of odd-numbered terms is not obvious; however, we believe it holds.   

BD give a slightly different argument for the vanishing of the $\Phi_m$ with $m$ odd.   Having calculated the Casimir term using the magnetic Green function, they calculate it again using the electric Green function, and find that the odd terms appear with reversed sign.  However, these must agree, since knowledge of the magnetic fields completely determines the electric fields.  Thus the odd terms must vanish. 

The frequency derivative that appears in the definition of $\Phi_m$ gives $m$ terms that differ only in the labeling (and thus are numerically equal), in which one of the $\bf K$ factors is replaced by
\begin{equation}
\label{explicitderK}
d{\bf K}(\vec r_{1},\vec r_{2})/dy = 
- \frac {y\rho^2}{1+y\rho} {\bf K}(\vec r_{1},\vec r_{2})
\end{equation}

When $\omega$ has a large positive imaginary part, $\bf K$ is exponentially small except when $ \rho = |\vec r_{i} - \vec r_{j}|$ is small. These implies that for large $|\omega|$, $\Phi_{4}$ (and higher-order $\Phi_{m}$) vanishes for large $\omega$, while $\Phi_{2}$ takes on a constant value that can be calculated by the surface integral over a combination of the principle curvatures $\kappa_{i}$ of the surface (the explicit expression is given below).

According to (\ref{kacnumber}) and (\ref{nocontour}), the Kac number is given by
\begin{equation}
\label{Kacnumber2}
{\cal K} =  \Im \int_0^{\infty + i\epsilon} \frac {2}{\pi \omega} \Phi(\omega) d\omega
\end{equation}
The integrand is analytic in the upper half-plane, so we can choose the contour that goes from the origin along the imaginary $\omega$ axis, and then returns to the real axis along an arc at large $|\omega|$.  The integrand is real on the first part of the contour and thus makes no contribution; it is $\Phi(\infty)/\omega$ on the arc, which leads to the conclusion ${\cal K} = -\Phi(i \infty)$.  

It is significant that the integral exists, because this implies that we can take the limit $\Omega \rightarrow \infty$ in Eq.(\ref{modegenerate}): in the expansion of $S(\Omega)$ in powers of $\Omega$, the first three terms vanish.  In contrast, a scalar field theory would surely have a nonvanishing second term, implying a contribution to the surface tension coming from vacuum fluctuations of the field \cite{KZLS}. 

In the large $y$ limit the only contribution to (\ref{ExplicitK}) comes from small separations, so that $\cal K$ can be calculated from the curvature elements of the surface.  Explicitly,
\begin{equation}
\label{Kacnumberresult}
{\cal K} =
\frac {1}{128\pi} \int_{s} (3 (\kappa_{1} - \kappa_{2})^{2} + 8\kappa_{1} \kappa_{2}) d^{2} r
\end{equation}
The second term reduces to the Gauss-Bonnet integral, which is a topological invariant:
\begin{equation}
\label{GB}
\int \kappa_{1} \kappa_{2}  d^{2}r = 4\pi (1-g)
\end{equation}
where $g$ is the genus of the surface ($g=0$ for singly connected surfaces; $g=1$ for the torus). 
We observe that (\ref{Kacnumberresult}) is positive for all singly connected surfaces: the introduction of a conducting surface slightly increases the number of modes.  The expression is also positive for surfaces of genus 1, however
this includes zero-frequency modes that represent a magnetic field that threads the space, which properly do not belong in the Kac number.

The Casimir term can be evaluated using the same contour. Using (\ref{casimirnumber}) and (\ref{nocontour}),  
\begin{eqnarray}
\label{Casimirresult}
{\cal C} &=& \Im \int_0^{\infty + i\epsilon} \Phi(\omega) \frac {d \omega
}{\pi}
\nonumber
\\
&=& \Im \int_0^{\infty + i\epsilon} (\Phi(\omega) + {\cal K}) \frac {d\omega 
}{\pi}
\nonumber
\\
&=&  \int_0^{\infty} (\Phi(iyc) + {\cal K}) \frac{dy}{\pi}
\end{eqnarray}
The addition of $\cal K$ to the integrand does not change the integral along the real $\omega$ axis, because it is adding a real number and we only need the imaginary part; however, upon changing the contour to the imaginary $\omega$ (real $y$) axis, it eliminates the contribution from the arc at large $|\omega|$. Since $\Phi(iy)$ approaches $-{\cal K}$ at large $y$, the addition makes the remaining integral convergent. 

For $\omega = iyc$, the tensors $\bf K$ are real.  There are an even number of these in every term of $\Phi$, and they are decreasing functions of $y$.  Then it is expected that each of the $\Phi_{2m}$ will be negative valued.  However, the integral of $\Phi_{2} + {\cal K}$ is usually positive, with the consequence that the Casimir term will be positive.  The known exception to this rule are long cylinders, where the $y$ integral of $\Phi_{2} + {\cal K}$ is small.

By means of the argument just presented, BD have reduced the determination of the Kac number and the Casimir term to the evaluation of some multiple integrals over known integrands. In Section III we will use these to determine the properties of some smooth surfaces.  We note that the BD paper goes beyond these results to discuss many other things, including the case of finite temperature, surfaces with sharp creases, and to verify that the expressions are consistent with the results for the parallel planes, cylinder, and sphere geometries calculated by other approaches.

Routes to the calculation of the Casimir interaction between objects have
been given previously\cite{Lifshitz,DLP,KK,EGJK,RMJJ,SGJ}.  We believe these
all develop along the same lines, differing mainly in notation and in how
the authors propose to turn their formal results into numerical values.  We
note that the self-interaction case is slightly more complicated that the case
of interaction between separate objects in that it needs the subtraction of
${\cal K}$ that appears in Eq. (\ref{Casimirresult}).

\section{Numerical evaluation}

The evaluation of $\Phi_{m}$ (Eq.(\ref{defPhi1})) requires  $m$ integrals over the surface; the result (\ref{Casimirresult}) is the integral over $y$ of the sum of the $\Phi_{2m}$.  These integrals are sufficiently complicated to defy analytic treatment, which is undoubtedly why BD's method has attracted so little attention.  However, they are tractable as numerical integrals.

A surface integral of a function $F(\vec r)$ can be written as an integral over the solid angle and estimated ("Monte Carlo integration"\cite{NR}) as an average over a large set of unit vectors $\hat r_{i}$ chosen randomly with uniform distribution of orientations
\begin{eqnarray}
\label{defMC}
\int F(\vec r) d^{2} r &=& \int F(\hat r s(
\hat r) ) W(
\hat r ) d\Omega
\nonumber
\\
&\approx & \frac {1}{N} \sum_{i=1}^N 
F(\vec r_{i}) W(\hat r_{i}) 
4\pi
\end{eqnarray}
where $d\Omega$ represents an integral over solid angles and $W(\hat r_i)$ is the weight function that translates solid angle into surface area, given by
\begin{equation}
\label{defW}
W(\hat r) = \frac { r^2}{\hat n \cdot \hat r}
\end{equation}
Then we can use (\ref{defMC}) to calculate the area of the surface
\begin{equation}
\label{defarea}
{\cal A} = \int W(\hat r) \frac {d\Omega}{4\pi}
\end{equation} 
and the Kac number 
\begin{equation}
{\cal K} = \frac {1}{128\pi} \int_{s} (3 (\kappa_{1} - \kappa_{2})^{2} + 8\kappa_{1} \kappa_{2}) W(\hat r) \frac {d\Omega}{4\pi} .
\end{equation}

Evaluation of the $\Phi_{m}$ requires a little more care, however.  The integrands are exponentially small when $y\rho > 1$, and are strongly peaked near the cases where all the points are close to each other.  Evaluation of the $m$-fold surface integrals by choosing $m$ uncorrelated directions ${\hat r}_i$ will give many small values and a few very large ones; the average (\ref{defMC}) will not have good statistics, especially when $y$ is large.  

The resolution of this problem\cite{NR} is choose the random vectors ${\hat r}_i$ from a joint distribution that emphasizes the small separation case, with a corresponding weight function $W(\hat r)$.   This can be viewed as a change in variables in the integration.   For example, consider the integral
\begin{equation}
\label{example}
\int \frac {d\Omega}{\sqrt{1 -(\hat z \cdot \hat r)^2|}} = \int_0^{2\pi} d\phi \int_0^\pi \frac {\sin \theta \:d\theta} {\sqrt {1-\cos^2\theta}}.
\end{equation}
In the first form, the integrand becomes arbitrarily large, and evaluating it by choosing uniformly distributed $\hat r$ is not going to work well; in the second form randomly sampling $\phi$ and $\theta$ is not a uniform distribution on the sphere, but the integrand is bounded (now regarding the $\sin\theta$ to be part of the integrand).  
 
Consider first $\Phi_{2}$.  According to (\ref{defPhi1}), (\ref{explicitK}), and (\ref{explicitderK}) it has the form
\begin{equation}
\label{Phi2calc1}
\Phi_{2}(iy) = \int dS_{1} \int dS_{2} (1+y\rho) 
\frac {(\hat n_{1} \cdot \vec\rho)(\hat n_{2} \cdot \vec\rho)y^{2}}
{4 \pi^{2}  \rho^{4}} \exp(-2y\rho)
\end{equation}
The integrand appears to be divergent for $\rho \rightarrow 0$ (that is, when $\vec r_1$ and $\vec r_2$ are close to each other); however, in this limit $\vec \rho$ almost lies in the surface and thus is nearly perpendicular to the surface normals $\hat n_1$ and $\hat n_2$.  The result is that the integrand is bounded.    Calling the integrand $F(\hat r_{1}, \hat r_{2})$, the integral can be rewritten as an integral over solid angles
\begin{eqnarray}
\label{Phi2calc2}
\Phi_{2} &=& \int d\Omega_{1} \int d\Omega_{2} F(\hat r_{1}, \hat r_{2}) W(\hat r_{1}) W(\hat r_{2})
\nonumber
\\
&=& \int \int d\Omega_{1} F(1,2) W(1) W(2) \sin (\theta) d\theta d\phi
\end{eqnarray}
where in the second equation we have written $\hat r_{2}$ in terms of the spherical angles relative to the direction $\hat r_{1}$.  The second form of the integral suggests that we can evaluate it by averaging over $(\theta_i,\phi_i)$ randomly and uniformly distributed (thus making the factor $\sin \theta$ part of the integrand); however, the integrand is exponentially small for $\rho y >> 1$, so that only the region of small $\rho$ is important.  This is equivalent to the region of small $\theta$, since $\rho \approx  r_1 \theta$ in this limit.
Therefore we replace $\theta$ by a new variable $Q$ defined by
\begin{equation}
\label{defQ}
Q = \frac {\sin (\theta/2) (1 + 2y r_{1})}{1+2y r_{1} \sin(\theta/2)}
\end{equation}
which has the features that $\theta = 0$ corresponds to $Q=0$, $\theta=\pi$ corresponds to $Q=1$, and for $y r_{1} \gg 1$, $Q=1/2$ means $y r_1 \sin(\theta/2) \approx 1$ (and thus $\theta$  small and $y \rho \approx 1$).  It follows from these properties that changing variables from $\theta$ to $Q$ gives a new integral for which the integrand makes significant contribution over about half the range of integration.  Randomly sampling $Q$ and $\phi$ involves averaging many numbers of about the same size, where randomly samply $\theta$ and $\phi$ would be the average of a few very large numbers and many small ones.

Calculating the other $\Phi_{m}$ benefits from this same idea,  because in the large $y$ limit the contribution is dominated by the configurations for which all of the $\hat r_{i}$ are close to each other.  

\section{A model for the Casimir term}

The Casimir term is an interaction between different parts of a surface (rather than a sum of local contributions, like the Kac number).  The ratio $1/{\cal C}$ is a length which in some way characterizes the size of the object, but it does not seem to be related to the obvious metrics, such as the ratio of volume to area or the root-mean-square separation of points on the surface or the surface integral of the average curvature. We give examples for various shapes in the next section; we find that the leading term ${\cal C}_2$ usually is the most important contribution, so that the Casimir term is almost the result of a pairwise interaction between points on the object.  The explicit form for the Casimir term given by BD is sufficiently complicated that it is difficult to anticipate what features of a surface contribute to it.  This led us to construct a geometric quantity $\cal G$ that is similar to $\cal C$ over a wide range of geometries.  

The effect of the tensor ${\bf K}$ (\ref{defK}) vanishes when the surfaces at $\vec R$ and $\vec r$ belong to the same plane, because in this case the field created by the current at $\vec r$ is perpendicular to the surface at $\vec R$ and thus induces no surface current in response.  This counteracts the large interaction between nearby points that might be expected.  These observations suggest that curvature of the surface is important to the Casimir term.  

We have constructed a representation capturing these properties in the form of a double integral over the surface of an integrand that depends on the separation $\rho$ of point pairs. To get the right scaling with size, the integrand must scale as $\rho^{-5}$.   The example of (\ref{Phi2calc1}) suggests one way to avoid a divergence of the integral at small $\rho$ is to have a numerator with four powers of factors such as $\hat n \cdot \hat \rho$.  Symmetry and the requirement that parallel planes attract leave few choices, and the only one that works at all well is
\begin{equation}
\label{geometrything}
{\cal G} =-  0.007 \int d^2 r \int d^2 r' (\hat n \cdot \hat \rho)
(\hat n' \cdot \hat \rho) (\hat n \cdot \hat n')
\frac {(\hat n \cdot \hat \rho)^2 + (\hat n' \cdot \hat \rho)^2}{\rho^5}
\end{equation}
The prefactor has been chosen to match the numerical results for $\cal C$ in the case of a sphere of radius $R$: ${\cal G} = 0.046/R= {\cal C}$.  But it works reasonably well for other geometries: for parallel planes with separation $D$ the value of ${\cal G}/{\cal A} = -0.0126/D^3$, to be compared with the Casimir interaction ${\cal C}/{\cal A} = -\pi^2/720 D^3 = -0.0137/D^3$. For the infinite cylinder, the value of ${\cal G}$ per unit length is zero (as is ${\cal C}_2$), while the Casimir term per unit length is small and negative. 

We ascribe no physical meaning to $\cal G$ beyond that just given; it is just a geometrical quantity.  However, it is a lot easier to calculate than $\cal C$. 

\section{Results}

In this Section we report the numerical results for some simple shapes:   ellipsoid of revolution, circular cylinder with hemispherical caps, torus, a "drum" (parallel circular disks joined with part of a torus), and a "cube" with rounded corners and edges.

For comparison please recall that for a sphere of radius $R$ the Kac number is $0.25$ and the Casimir term is $0.04618/R$, and for an infinite cylinder of radius $R$ the Kac number per unit length is $3/(64R) = 0.0469/R$ and the Casimir term per unit length is $-0.01356/R^{2}$.  The interaction energy per unit area of parallel planes with separation $D$ is given by a Casimir term $-\pi^2/720D^3 = -0.014/D^3$.

We believe the numerical values to be accurate to the number of places given.  In most cases ${\cal C}_{2}$ is the dominant term, though we note again that for the infinite cylinder ${\cal C}_{2}/L = 0$, so that the higher order terms are necessarily relevant.  We have calculated ${\cal C}_2$, ${\cal C}_4$, and ${\cal C}_6$ for all geometries, but supressed the reporting of ${\cal C}_6$ when it turned out to be negligibly small.

For the case of the sphere, we find ${\cal C} = {\cal C}_{2} + {\cal C}_{4} + {\cal C}_{6} = 0.0497 - 0.0008 - 0.0000 = 0.048 \pm 0.001$.   Although this agrees well with the accepted value, we note that it is also an easier calculation than most that we report, since the radius and curvature are constant.

\subsection{Cylinder with spherical caps}

The Casimir energy per unit length for an infinite cylinder is negative, while the Casimir energy for the sphere is positive.  As a way of seeing how the two limits are connected, we considered a cylinder of length $b$ and unit radius, with hemispherical caps on the ends.  For this object the area is given by
\begin{equation}
\label{cylinderarea}
{\cal A} = 2\pi (2 + b)
\end{equation}
and the Kac number (\ref{Kacnumberresult}) is
\begin{equation}
{\cal K} = \frac {16+3b}{64}
\end{equation}
We considered various values for $b$, as listed in the tablee below.

\begin{tabular} {c c c c c c c}
b & ${\cal A}$ & ${\cal K }$ & ${\cal C}_{2}$ & ${\cal C}_{4}$ & ${\cal C}$ & ${\cal G}$\\
\hline
0.1 & 13.19 & 0.254 & 0.0557 & -0.0008 & 0.055 & 0.046\\
0.2 & 13.82 & 0.259 & 0.0586 & -0.0009& 0.058 & 0.047\\
0.4 & 15.08 & 0.269 & 0.0602 & -0.0010& 0.059 & 0.048\\
0.6 & 16.34 & 0.278 & 0.0612 & -0.0013& 0.060 & 0.049\\
0.8 & 17.60 & 0.288 & 0.0623 & -0.0016 & 0.061 & 0.049\\
1.0 & 18.85 & 0.297 & 0.0627 & -0.0019 & 0.061& 0.050\\
2.0 & 25.14 & 0.344 & 0.0645 & -0.0035 & 0.061& 0.050\\
4.0 & 37.71 & 0.438 & 0.0669 & -0.0070 & 0.060& 0.051\\
8.0 & 62.78 & 0.624 & 0.0700 & -0.014 &0.056 & 0.050 \\
\hline
\end{tabular}
 
This table can be simply summarized: $\cal C$ is small and nearly constant over the range of $b$ considered; ${\cal C}_2$ is the dominant contribution to $\cal C$; and $\cal G$ is a good approximation to $\cal C$.

The case $b=0$ is the sphere, and the case of an infinite cylinder is relevant to the large $b$ limit.  For large $b$, ${\cal C}_2$ is only growly slowly, if at all, while ${\cal C}_4$ is negative and proportional to $b$. This is consistent with what is known about the infinite cylinder, for which the Casimir term per unit length is negative.  For very large $b$, $\cal C$ is expected to go negative due to ${\cal C}_4$ and higher order terms, but this is not apparent from the table.  ${\cal C}_2$ and $\cal G$ have a finite positive value from the ends, and get only a small contribution from the cylindrical part of the object; it will be a positive quantity for all $b$. 

\subsection{Torus}

The torus is described in parametric form by
\begin{eqnarray}
\label{deftorus}
x = (R + \cos \theta) \cos \phi
\nonumber
\\
y = (R + \cos \theta) \sin \phi
\nonumber
\\
z = \sin \theta
\end{eqnarray}
where the major radius $R$ has to be larger than the minor radius  (which has been chosen to be unity) and greater than $2$ if we are to avoid regions with large curvature.   The area is ${\cal A} = 4\pi^2 R$ and the value of (\ref{Kacnumberresult}) is
\begin{equation}
\label{toruskac}
{\cal K} = \frac {3 \pi R^2}{32 \sqrt{R^2-1}}
\end{equation}
BD note that this counts a static magnetic field as a mode, which in some contexts is incorrect.   In comparing the results to those for other shapes, it should be noted that the distance around the torus is approximately $L = 2\pi R$, and that it is this dimension that best corresponds to the parameter $b$ for the cylinder and ellipsoid.  The Kac number is minimum ($3\pi/16$) for $R=\sqrt 2$.

\begin{tabular} {c c c c c c c c}
R & ${\cal A}$ & ${\cal K }$ & ${\cal C}_{2}$ & ${\cal C}_{4}$ & ${\cal C}_{6}$ &${\cal C}$ & ${\cal G}$\\
\hline

2. &  79.  &  0.68  &  0.040 & -0.112 & -0.040 & -0.11 & 0.025 \\
3. & 118.  &  0.94  &  0.031 & -0.146 & -0.047& -0.16 & 0.017\\
4. & 158.  &  1.217  &  0.029 & -0.188 & -0.060 & -0.22 & 0.014\\
5. & 197. &  1.50  &  0.028 & -0.232 & -0.073 & -0.28 & 0.010\\
6. & 237.  &  1.79  &  0.029 & -0.275 & -0.087 & -0.33 & 0.009\\
8. & 316.  &  2.37  &  0.035 & -0.364 & -0.116 & -0.44 & 0.007\\
10. & 395. &  2.96  &  0.040 & -0.455 & -0.141 & -0.56 & 0.006\\
\hline
\end{tabular}

\begin{figure}
\includegraphics[width=1.0\columnwidth,keepaspectratio]{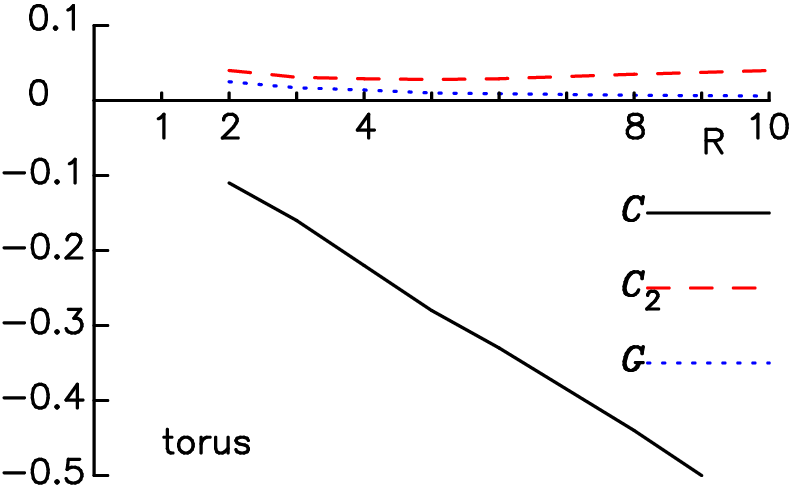} 
\caption{(color online) The Casimir term for the torus, as a function of its major radius $R$ (the minor radius is unity).
}
\end{figure} 
The torus is similar to the cylinder, especially when R is large.  Accordingly, the Casimir term is negative, and nearly proportional to the circumference: ${\cal C} \approx -0.009 \times 2 \pi R$.  The leading term ${\cal C}_2$ increases with $R$, but not as fast as the length itself.  The function $\cal G$ is small and positive and similar to ${\cal C}_2$, but these are not good approximations to $\cal C$.  

\subsection{Drum}

This object consists of a pair of parallel disks of radius $R$, joined at their boundaries by the external part of a torus with unit minor radius.   For $R$ = 0 it is a sphere of unit radius, while for large $R$ it approximates parallel planes with separation $2$.

\begin{tabular} {c c c c c c c}
R & ${\cal A}$ & ${\cal K }$ & ${\cal C}_{2}$ & ${\cal C}_{4}$ & ${\cal C}$ & ${\cal G}$\\
\hline
0.05  & 13.6  &  0.2509  &  0.0487 & -0.0008 & 0.048 & 0.045\\
0.2 &  16.8 &   0.2588  &  0.0488 & -0.0008 & 0.048 &0.043\\
0.3  & 19.  &  0.2666   & 0.0495 & -0.0009 & 0.048 & 0.042\\
0.5  & 24.  &  0.2856  &  0.0512 & -0.0012& 0.050 & 0.042\\
1. &  39. &   0.3438 &   0.0554 & -0.0021& 0.053 & 0.039\\
2. &  77. &   0.4767 &   0.0586 & -0.0045 & 0.054 & 0.021\\
3. & 128. &   0.6173 &   0.0555 & -0.0076 & 0.057 & -0.015\\
4. & 192  &   0.7598 &   0.041 & -0.011 & 0.040 & -0.07\\
5. & 268 &   0.9062 &   0.016 & -0.015 & 0.001 & -0.15\\
6. & 357. &   1.0509 &  -0.018 &-0.019 & -0.037 & -0.24\\
8. & 572. &   1.3417 &  -0.117 & -0.027& -0.144 & -0.49\\ 
10. & 839. &    1.6375 &   -0.244 & -0.036& -0.28 & -0.82\\
\hline
\end{tabular}

\begin{figure}
\includegraphics[width=1.0\columnwidth,keepaspectratio]{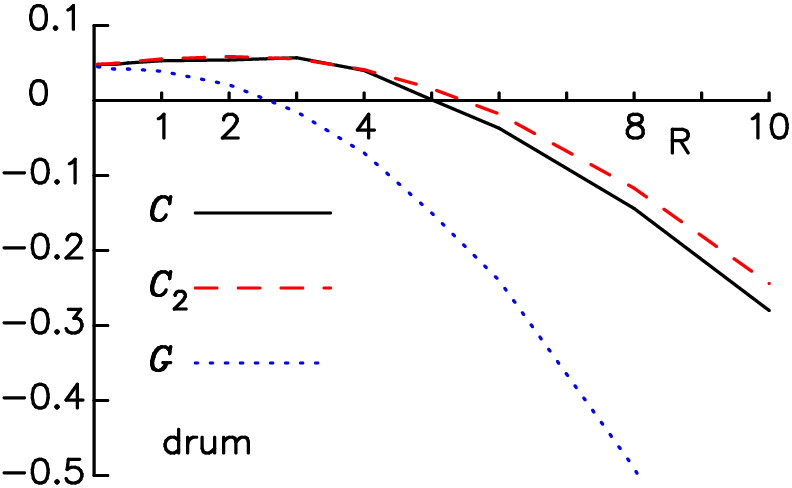} 
\caption{(color online) The Casimir term for the drum of radius $R$ and height $2$.  For $R$ less than 3, the curved region is dominating the energy, which increases because the area of the curved region is increasing.  For large $b$ the attraction between the parallel faces is more important, with the result that the Casimir term decreases and goes negative. For $R = 10$ the Casimir term is approximately that of a pair of parallel disks with area $\pi R^2$, which would give ${\cal C}_{approx} = -\pi^3 R^2/5760 = -0.54$.
}
\end{figure} 
 
\subsection{Ellipsoid of revolution}

We defined the ellipsoid of revolution by the condition
\begin{equation}
\label{defellipsoid}
 x^2 + y^2 + z^2/b^2 = 1
\end{equation}
where the ellipsoid has semi-principal axes $b$, $1$, and $1$. 

\begin{tabular}{c c c c c c c c}
b & ${\cal A}$ & ${\cal K }$ & ${\cal C}_{2}$ & ${\cal C}_{4}$ & ${\cal C}_{6}$ &${\cal C}$ & ${\cal G}$\\
\hline
0.2 & 6.87 & 1.65 & 1.26 & -0.076 & -0.004& 1.18 & 2.02\\
0.3 & 7.39 & 0.805 & 0.414 & -0.0815 & -0.001& 0.33 & 0.33   \\
0.4 & 8.00 & 0.51 & 0.196 & -0.013 & -0.0004 & 0.18 & 0.24\\
0.5 & 8.67 & 0.379 & 0.118 & -0.0211 & -0.0002 & 0.096 &0.13\\
0.6 & 9.39 & 0.314 & 0.083 & -0.003 & -0.0001 & 0.08 & 0.087 \\
0.8 & 10.93 & 0.261 & 0.057 & -0.001 &0 & 0.056 & 0.054\\
1.0 & 12.57 & 0.250 & 0.0497 & -0.001 & 0. & 0.049 & 0.046\\
1.5 & 16.9 & 0.280 & 0.0527 & -0.002 & -0.000 & 0.051 & 0.050\\
2.0 & 21.5 & 0.336 & 0.062 & -0.004 & -0.0001& 0.058 & 0.061\\
2.5 & 26.1 & 0.400 & 0.073 & -0.007 & -0.0039 & 0.062 & 0.075 \\
3.0 & 30.9 & 0.467 & 0.085 & -0.007 & -0.0005 & 0.07 & 0.090\\
4.0 & 40.5 & 0.609 & 0.115 & -0.017 & -0.0100 & 0.083 & 0.12\\
6.0 & 60. & 0.899 & 0.175 & -0.032 & -0.0023 &0.14  & 0.18\\
8.0 & 79. & 1.186 & 0.215 & -0.050 & -0.0045 &0.16 & 0.26\\
10.0 & 99. & 1.49 & 0.29 & -0.075  & -0.004 & 0.21  &0.32\\

\hline
\end{tabular}

\begin{figure}
\includegraphics[width=1.0\columnwidth,keepaspectratio]{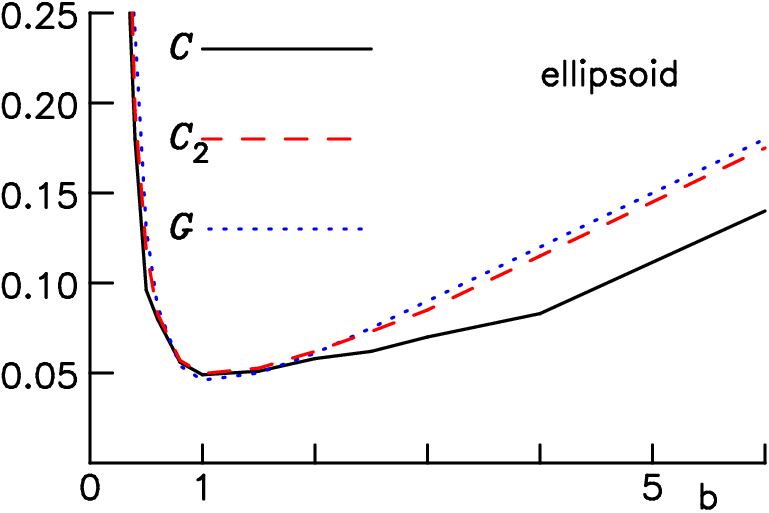} 
\caption{(color online) The Casimir term for the ellipsoid with semiaxes $b,1,1$.  For large $b$ the Casimir term per unit length is comparable to that for an infinite cylinder. However, for small $b$ (the limit of extremely oblate ellipsoids), the energy is diverging positive, rather than going negative:  apparently the curvature of the rim of the ellipsoidal disk is a more important effect than the attraction between the two nearly parallel faces.  }
\end{figure} 

The minimum value for the Casimir term with unit transverse axis length occurs for the slightly prolate ellipsoid $b = 1.1 $.  However, if we fix the area or volume, the minimum value is for the sphere ($b=1$) to the accuracy that we can determine this.  
 
\subsection{Cube with rounded corners and edges}

For large even $m$, the surface
\begin{equation}
\label{defcube}
 x^m + y^m + z^m = 1
\end{equation}
is approximately the unit cube, but with rounded edges and corners.
 
\begin{tabular} {c c c c c c c}
m & ${\cal A}$ & ${\cal K }$ & ${\cal C}_{2}$ & ${\cal C}_{4}$  &${\cal C}$ & ${\cal G}$\\
\hline
2 & 12.57 & 0.25 & 0.0487 & -0.0008   &0.048 & 0.045\\
4 & 17.59 & 0.43 & 0.17 & -0.004 & 0.16  & 0.178\\
6 & 19.62 & 0.66 & 0.35 & -0.010 & 0.34& 0.43\\
8 & 20.68 & 0.89 & 0.57 & -0.015 & 0.55& 0.80\\
10 & 21.34 & 1.13 & 0.86 &-0.021  & 0.84 & 1.29\\
12 & 21.78 & 1.37 & 1.14 &-0.026  &  1.11& 1.87\\
\hline
\end{tabular}

For large $m$, the area approaches $24$.  The results suggest that in the limit of large $m$, $\cal K$ diverges proportional to $m$, while $\cal C$ diverges proportional to $m^2$.  This conclusion is not in conflict with the calculation by Lukosz \cite{lukosz}, because he only considered the modes in the interior of a cube.
\begin{figure}
\includegraphics[width=1.0\columnwidth,keepaspectratio]{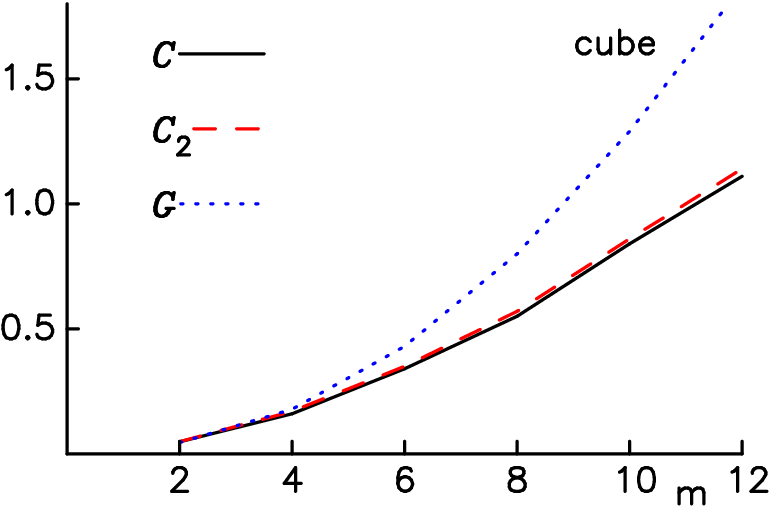} 
\caption{(color online) The Casimir term for the soft-edged cube, as a function of the exponent $m$
}
\end{figure} 

\subsection{Summary}

 For most geometries, the value of the Casimir term is largely due to the first term ${\cal C}_2$, and then $\cal G$ is also qualitatively accurate.  The important exception is the cylinder and cylinder-like objects, for which ${\cal C}_2$ is small and the higher order terms play a more important role.  We can very roughly summarize our results by saying that the parts of a surface of low curvature have small self-energy; curved surfaces have a positive self-interaction; parts of a surface that are roughly parallel with small separation give a negative contribution, and the interaction between distant parts of a surface do not give a significant contribution to the Casimir term.
Our model function reproduces these features.

\section{Acknowledgements}

This work was supported by US AFOSR Grant No. FA9550-11-1-0297.


\begin{thebibliography} {10}
\bibitem{note1}
For an infinite region, the mode frequencies become dense and the poles of $\Gamma$ merge into a branch cut where $\Gamma$ is discontinuous.  It is conceptually simpler to consider the effect of the introduction of a conducting surface into a large but finite region of space, and then take the limit of an infinite region later.  However, we point out an essential role played by the cutoff:  so long as $\Omega$ is finite, we can ignore the region of very large real $\omega$ without much concern for the behavior of the integrand there (it amounts to the neglect of the large $\omega_\alpha$ part of Eq.(\ref{defmodesum})).  Taking the limit $\Omega \rightarrow \infty$ before the integration is done leaves us with an integral that is not well defined.
\end{thebibliography}
\end{document}